%% file: main.tex
\documentclass[reprint,superscriptaddress,floatfix,preprintnumbers,prl,nofootinbib]{revtex4-2}

\usepackage[colorlinks=true,breaklinks=true]{hyperref}
\usepackage[utf8]{inputenc}
\hypersetup{allcolors=[rgb]{
0.0 0.0 0.6},linkcolor=[rgb]{0.75 0.05 0.05}}
\usepackage{mathtools}
\usepackage{xspace}
\usepackage{graphicx}
\newcommand{\unit}[1]{\ensuremath{\mathrm{\,#1}}\xspace}

\newcommand{\e}{\unit{e^{-}}}

\usepackage{amsmath}
\usepackage[utf8]{inputenc}
\usepackage[T1]{fontenc}
\usepackage{textcomp}
\usepackage{gensymb}
\usepackage{newunicodechar}
\newunicodechar{°}{\degree}
\usepackage[dvipsnames]{xcolor}
\usepackage{float}
\usepackage{amsmath}
\usepackage{empheq}
\usepackage[theorems,skins]{tcolorbox}
\usepackage{letltxmacro}
\LetLtxMacro{\oldcite}{\cite}
\usepackage{cancel}
\renewcommand{\cite}[1]{\mbox{\oldcite{#1}}}
\newtcolorbox{mymathbox}[1][]{colback=white, sharp corners, #1}
\usepackage{orcidlink}
\usepackage{xcolor}
\usepackage[modulo]{lineno}

\long\def\exclude#1{}
\newcommand{\beq}{\begin{equation}}
\newcommand{\eeq}{\end{equation}}

\newcommand{\bs}[1]{\boldsymbol{#1}}

\def\ga{\,\,\raise0.14em\hbox{$>$}\kern-0.76em\lower0.28em\hbox
{$\sim$}\,\,}

\allowdisplaybreaks
\usepackage{amsmath}
\usepackage{dsfont}

\usepackage{braket}

\begin{document}

\title{SENSEI: A Search for Diurnal Modulation in sub-GeV Dark Matter Scattering}

\include{authors}

\date{\today}

\begin{linenumbers}
\begin{abstract}
Dark matter particles with sufficiently large interactions with ordinary matter can scatter in the Earth’s atmosphere and crust before reaching an underground detector. This Earth-shielding effect can induce a directional dependence in the dark matter flux, leading to a sidereal daily modulation in the signal rate. We perform a search for such a modulation using data from the SENSEI experiment, targeting MeV-scale dark matter. 
We achieve an order-of-magnitude improvement in sensitivity over previous direct-detection bounds for dark-matter masses below $\sim 1~\mathrm{MeV}$, assuming the Standard Halo Model with a Maxwell--Boltzmann velocity distribution, and restrict the amplitude of a general daily modulation signal to be below 6.8 electrons/gram/day. 
\end{abstract}
\end{linenumbers}

\maketitle

\section{Introduction} 
There is substantial evidence that a large fraction of the matter in our universe consists of a non-baryonic form of matter known as dark matter (DM). Despite direct-detection experiments searching for DM for the past four decades, no convincing evidence for DM has been found. DM candidates with mass less than the proton (``sub-GeV DM'') are well-motivated and have been given much attention over the last decade~\cite{battaglieri2017cosmicvisionsnewideas,essig2023snowmass2021cosmicfrontierlandscape}. DM particles in this mass range can scatter off electrons and deposit energies of $\mathcal{O}$(few~eV)~\cite{Essig:2011nj, Essig:2012yx}. Consequently, semiconductor-based detectors with eV-scale ionization thresholds are well-suited for these searches~\cite{Essig:2011nj,Essig:2015cda,Graham:2012su}, providing the sensitivity required to probe sub-GeV DM candidates.

Particle interactions in silicon can excite electrons from the valence band to the conduction band, leading to the creation of one or more electron-hole pairs (simply referred to as electrons, \e). The Skipper Charge-Coupled-Device (Skipper-CCD) is a silicon detector that can measure these signals~\cite{Tiffenberg_2017}. The charge in each pixel of a Skipper-CCD is moved pixel-by-pixel to the readout stage, where it is measured repeatedly and non-destructively to achieve sub-electron noise precision. However, in the lowest charge bins, the detector sensitivity is limited by  backgrounds~\cite{sensei2020,Du_2022,Du:2023soy,Baxter:2025odk}. This makes it especially challenging to search for DM particles with masses near the MeV scale, as these would only populate the lowest charge bins. One method to substantially improve our sensitivity to such DM particles is to search for a time-varying signal over a time-independent background. 

There are several known reasons why the DM scattering rate might modulate.  This includes an annual modulation from the Earth moving around the Sun~\cite{drukier_1986}, a daily modulation from gravitational focusing~\cite{nielsen_grav_focusing,Sikivie_2002,Alenazi_2006}, and the small modulation induced by the Earth's velocity with respect to the galactic halo~\cite{nielsen_grav_focusing}.  In addition, the modulation can be induced by DM scattering in the Earth before interacting in the detector.  One example is DM that couples to the visible sector via a dark photon that is kinetically mixed with the ordinary photon, which predicts that DM couples both to electrons and nucleons. 
For sufficiently large DM interactions, scattering in the Earth distorts the DM velocity distribution before the DM reaches underground detectors~\cite{COLLAR1992181,PhysRevD.47.5238,Hasenbalg_1997,Emken_2017}. As a result, the observable DM flux becomes dependent on the detector’s position relative to the incoming DM wind, leading either to an attenuation or an enhancement of the detectable flux due to this anisotropic scattering effect. The resulting daily modulation has been recently studied theoretically~\cite{Bertou:2025adb} and  experimentally~\cite{LuxModulationSearch,Arnquist_2024}. 
In this \textit{Letter} we present the first results from a daily modulation search with the SENSEI experiment using data taken at the underground MINOS cavern at Fermi National Accelerator Laboratory (Fermilab). We perform both a model-independent and a model-dependent search, present a new modulation analysis tool, and demonstrate an improved limit for DM masses $m_\chi < 2$~MeV.

Supplemental Materials provide additional analysis details. 

\section{Data Collection}
The dataset used here—collected at MINOS between 29 Feb and 16 May 2024—was previously described in Ref.~\cite{sensei1epaper}, where it was used to characterize the detector’s overall single-electron rate by comparison with that study’s primary SNOLAB dataset. Data were collected using two science-grade Skipper-CCDs designed at LBNL and fabricated at Teledyne DALSA Semiconductor, each of which has a thickness of 665 $\mu$m and an active area of $6144 \times 1024$ pixels of size $15 \times 15 \,\mu \text{m}^2$, for an active mass of $2.19 \,\text{g}$. 
Each CCD is read out through four amplifiers, each of which reads a ``quadrant'' of 512 rows by 3072 columns. 

We refer to a single exposure and readout of both Skipper-CCDs, with a total of 8 quadrants, as an ``image.''
To mitigate the effects of exposure-independent charge accumulation in the serial register, the data are binned during readout by summing 32 rows into the serial register at a time. As a result, each “superpixel” in the image corresponds to a $1 \times 32$ block of physical pixels. 
With four amplifiers, each reading one CCD quadrant including $128\times4$ superpixels of overscan per quadrant, the total readout time per image is approximately 16 minutes.

While the data from MINOS described in~\cite{sensei1epaper} include cyclic exposures of 0, 2, and 6 hours, our analysis is restricted to the 6-hour exposures. 
By refraining from any prior daily-modulation analysis of this dataset, we ensure our search remains unbiased. To ensure a hidden-data analysis, we develop and validate the analysis framework on two sets of three consecutive days, amounting to 12 images, which represent approximately 8\% of the full dataset. We then perform the analysis on 141 six-hour images (comprising the DM search dataset), collected over a period of 75 days, with a quadrant-dependent $1e^-$ density ranging from $1.09 \times 10^{-5}$ $e^-/\mathrm{pixel}$ to $2.46 \times 10^{-5}$ $e^-/\mathrm{pixel}$.

\section{Data Reconstruction \& Selection}

Data processing, calibration, and corrections are done using the exact same procedure as in~\cite{sensei1epaper}. The masks used are also identical as those described in~\cite{sensei1epaper} with the caveat that the hot image mask is not used as it could introduce/hide a modulation signal. The noisy-image mask is applied, although we find zero ``noisy'' images. The masks remove $\approx 23\%$ of the pixels across all images. 

\section{Signal Modeling}
\subsection{Model-Independent Rate}

For a model-independent signal, we fit the number of events sourced by the DM to a constant that includes both the background and an average DM signal rate, together with a single sine function with a period of one day, defined as
\begin{equation}
\label{eq:mod_ind_signal}
    S(\mu) = \int_{t_{exp}} a_1 \cdot \sin(2\pi\omega t + \phi)dt    \,.
\end{equation}
Here $a_1$ is the amplitude of the signal's daily modulation rate, $t_{exp}$ is the exposure duration, and $\omega = \frac{1}{day}$ is the angular frequency corresponding to a daily modulation rate. Consequently, we define the model-independent signal parameters as $\mu = (a_1, \phi)$. The overall rate is constrained to be higher than 0 at all times.

\subsection{Model-Dependent Rate} 

Within specific models the scattering rate and modulation amplitude can be calculated as a function of the model parameters.  
For concreteness, we consider here the case of sub-GeV DM interacting with the visible sector through a dark photon that is kinetically mixed with the ordinary photon~\cite{Holdom:1985ag,Galison:1983pa,Essig:2011nj,Essig:2015cda,essig2023snowmass2021cosmicfrontierlandscape}.  Such DM particles can scatter off both nuclei and electrons, with the former dominating the scattering in the atmosphere and the Earth, and the latter creating the observable signal in the detector~\cite{Emken:2019tni}. 

The DM-electron scattering rate is given by~\cite{Essig:2015cda} 
\begin{equation}
\frac{\mathrm{d}R}{\mathrm{d}\ln{E_e}} = \frac{\rho_\chi}{m_\chi} \frac{\bar{\sigma}_e}{8 \mu_{\chi e}^2} \int \frac{\mathrm{d}q}{q^2} \left| F_{\text{DM}}(q) \right|^2 \left| f_{\text{res}}(E_e,q) \right|^2  \eta(v_{\text{min}})\,.
\label{eq:model_dep_model}
\end{equation}
Here, $E_e$ and $q$ denote the energy transferred to the electron and the momentum lost by the DM, respectively. The DM mass is denoted by $m_\chi$, and the local DM density is taken to be $\rho_\chi = 0.3\, \text{GeV}/\text{cm}^3$ recommended from~\cite{Baxter_2021} (see also~\cite{Catena:2009mf, Salucci:2010qr, Read:2014qva, deSalas:2020hbh,Lim:2023lss}). The quantity $\mu_{\chi e}$ is the reduced mass of the DM–electron system, and $\bar{\sigma}_e$ is the reference DM–free-electron scattering cross section, defined at a fixed momentum transfer $q_0 = \alpha m_e$, with $m_e$ the electron mass.
The function $F_{\text{DM}}(q) = (\alpha m_e / q)^n$ encapsulates the momentum dependence of the interaction. Specifically, $n = 0$ corresponds to the case of a heavy mediator ($m_{A'} \gg \alpha m_e$, where $m_{A'}$ is the dark-photon mass), while $n = 2$ describes the light mediator limit ($m_{A'} \ll \alpha m_e$).
The material-specific, dimensionless response function $f_{\text{res}}(E_e, q)$ characterizes the probability of an electron excitation with energy $E_e$ and momentum transfer $q$ in the target~\cite{Essig:2015cda}.

Crucially for the present analysis, the dependence on the DM halo velocity distribution is encoded in 
\begin{equation}
\eta(v_{\text{min}}) =  \int_{v_{\text{min}}} \frac{f(\mathbf{v}, t)}{v} \, \mathrm{d}^3\mathbf{v}\,.
\end{equation} 
The DM velocity distribution is typically modeled as an isotropic Maxwell–Boltzmann distribution in the Galactic frame~\cite{Green_2017}. In this work, we adopt the Standard Halo Model with parameters $v_0 = 238$~km/s and $v_{\text{esc}} = 544$~km/s, following~\cite{Baxter_2021}. The Earth's velocity is computed as a function of time and evaluated on March 3rd, 2024, corresponding to the midpoint of the data-taking period.
However, this distribution is subsequently modified as DM particles traverse the Earth.
We define the isodetection angle (or isoangle) $\Theta$ as the angle between the local zenith and the direction of the mean DM flux, given by $\braket{\vec{v}_\chi} = -\vec{v}_{\text{lab}}$. Here, $\Theta = 0^\circ$ corresponds to a flux arriving from directly overhead, while $\Theta = 180^\circ$ corresponds to a flux arriving from directly beneath the detector.
In most DM searches, the directionality of the flux is neglected. However, if the DM–nucleus scattering cross section is sufficiently large, DM particles may undergo scattering within the Earth’s interior---particularly in the mantle and core---before reaching the detector. This can significantly alter the incoming DM flux and the observable signal.
Importantly, the isoangle $\Theta$ varies with time due to the Earth's rotation. Over the course of a sidereal day, $\Theta(t)$ modulates at any fixed detector location. For our experimental setup at Fermilab ($41.84^\circ$~N), $\Theta(t)$ varies between $10^\circ$ and $86^\circ$ (for more details see~\cite{Emken_2017, Emken:2019hgy, Bertou:2025adb}). Consequently, for models with sufficiently large DM–nucleus cross sections, a characteristic daily modulation of the signal is expected.

To quantify this effect, we employ the {\tt DaMaSCUS} code~\cite{Emken_2017}, which performs a full 3D Monte Carlo simulation of DM propagation through the Earth, accounting for energy loss and deflection due to elastic scattering. 
The resulting distribution, $f(v, \Theta_i)$, can be integrated and convolved with the relevant detector response to obtain the differential event rate as a function of isodetection angle. Since $\Theta_i$ is a deterministic function of time at a fixed location, it is straightforward to convert $\Theta_i \to t_i$, allowing one to express the event rate as a function of time. By implementing the ionization model from~\cite{RamanathanKurinskyIonization}, and using a modified version of {\tt QCDark}~\cite{dreyer2023fullyabinitioallelectroncalculation}, which incorporates a charge-screening function based on the formalism in~\cite{PhysRevB.47.9892} (for details see also~\cite{Bertou:2025adb}),
we compute the number of time-dependent single-electron events from the rate as:

\begin{equation}
\label{eq:mod_dep_sig}
S({{\bf \mu}}) = \int_{t_{exp}} R_{1e^-}(m_\chi, \bar{\sigma}_e,t)\, dt,    
\end{equation}
where ${\bf \mu} = (m_\chi, \bar{\sigma}_e,t)$, and $t_{exp}$ is the exposure duration. 

\section{Statistical Treatment}
\label{sec:modulation_analysis_main}

Using the number of unmasked pixels containing 1\e events, we perform a likelihood ratio analysis based on~\cite{cowan2011asymptotic} to search for a potential DM signal. Our analysis framework is general and accommodates both model-independent (Eq.~\eqref{eq:mod_ind_signal}) and model-dependent (Eq.~\eqref{eq:mod_dep_sig}) signals. 

To search for a DM signal, we compute the profile likelihood–ratio test statistic based on the background-only hypothesis, $\mu=0$.
We then compute a $p\text{-value}$ as the percentile of the observed test statistic within the distribution of toy Monte Carlo simulations under the background-only hypothesis.
In case no evidence for a DM signal is found, we set a 90\% C.L.\ upper limit on the signal strength $\mu$. For each $\mu$, we compute the statistic of the observed data, simulate its distribution under the signal-plus-background hypothesis with toy Monte Carlo, and identify the $\mu$ at which the observed test statistic reaches the 90th percentile. The full details of the likelihood analysis computations and the Monte Carlo simulation are provided in the Supplemental Materials.

In the single-image likelihood function used in this analysis, we assume Poisson statistics with a constant dark current across all quadrants throughout data collection. Upon inspecting the data, we observed deviations from a pure Poisson distribution. To ensure these deviations do not bias the results, we performed a dedicated validation test (see Supplementary Materials).
Furthermore, we conducted an additional test to verify that the time-independent background component is consistent with Poisson statistics under the test statistic employed in our analysis. To evaluate this consistency, we applied the bootstrap method~\cite{Efron:1979, EfronTibshirani:1993}, which involves uniformly resampling the \(1\e\) time-ordered dataset to produce new datasets with randomly shuffled timestamps. This procedure effectively removes any time-dependent structure, allowing us to isolate and test the time-independent background. It should be noted, however, that this method cannot detect small deviations from Poisson statistics that arise from time-independent sources.

\begin{figure}[!t]
\centering
\includegraphics[scale=0.45]{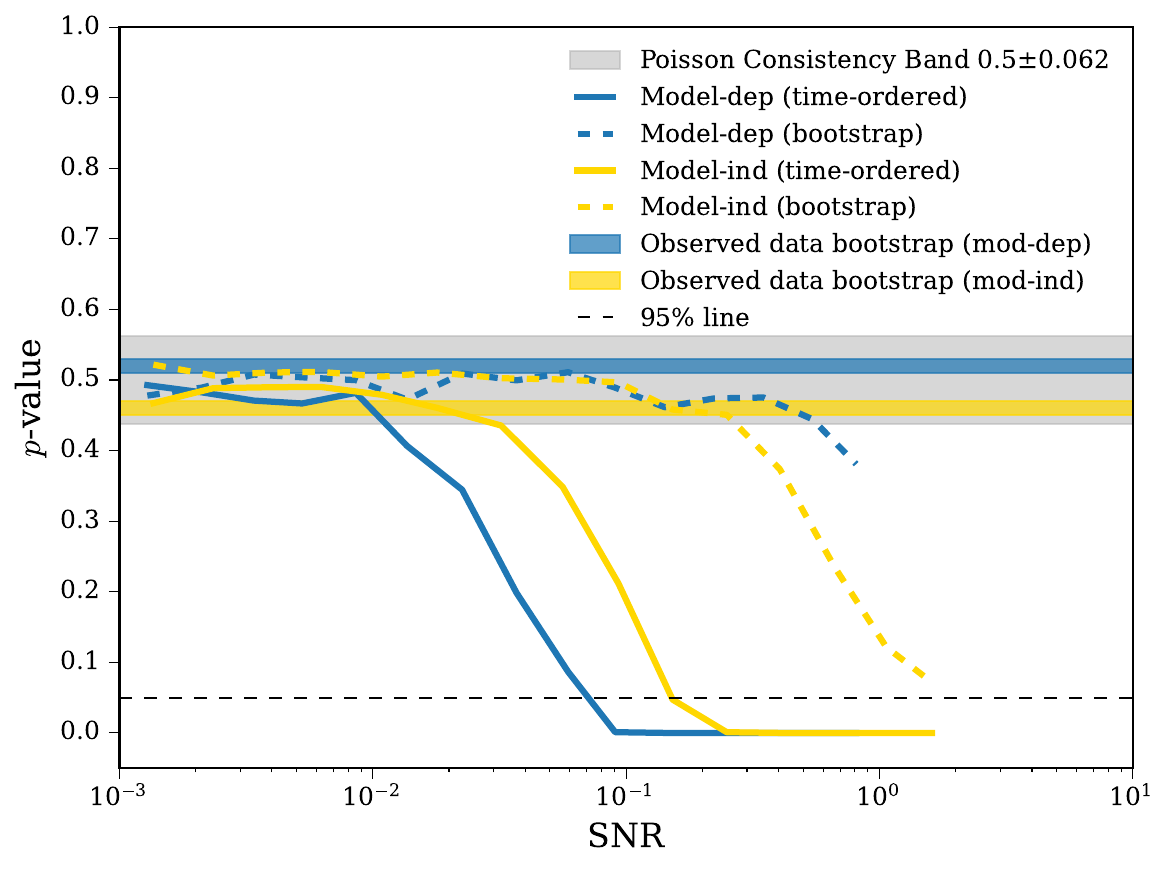}
\caption{Results of the bootstrap test validating the assumption of a Poisson-distributed, time-independent background of our observed data. The \textbf{gray band} shows the 2$\sigma$ variance around a mean $p$-value calculated from fitting background-only Poisson based Monte Carlo datasets. {\bf The blue and yellow bands} indicate the calculated variance of the $p$-value obtained from bootstrapping the observed data multiple times. No statistical deviation of the background from the Poisson assumption is observed. All horizontal bands are signal-to-noise ratio(SNR) independent. For the simulation bootstrap test the $p$-value is shown as a function of the SNR. For each value of SNR, the mean $p$-value from 500 simulated datasets was calculated, assuming a model-independent (Eq.~\eqref{eq:mod_dep_sig}; {\bf yellow}) and model-dependent (Eq.~\eqref{eq:mod_ind_signal}; {\bf blue}) signal.  The {\bf solid lines} indicate fits using time-ordered timestamps, while {\bf dashed lines} indicate fits using bootstrap-resampled timestamps.}
\label{fig: boot_simulation}
\end{figure}

\begin{figure}[!t]
  \centering
  \includegraphics[width=\linewidth]{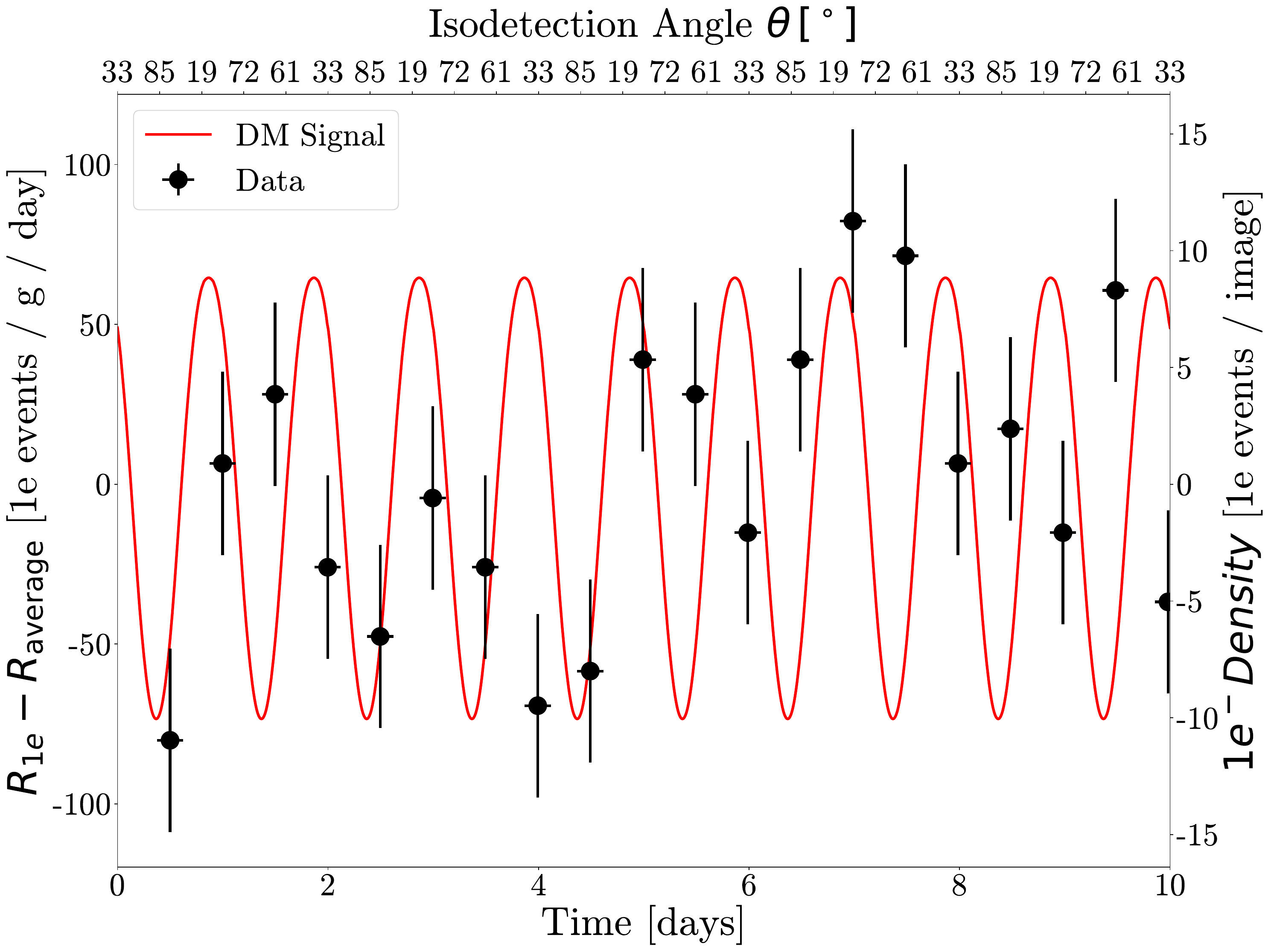}
  \caption{Comparison of the fluctuation around the average of both the measured 1\e rate ({\bf black circles}) with the rate predicted for a 1~MeV DM particle interacting with ordinary matter through a heavy dark-photon mediator with a cross section of
$\sigma_e = 5\times10^{-32},\mathrm{cm}^2$ ({\bf solid red line}), which represents the current most stringent bound at this mass. The measured rate is taken from the lowest-noise CCD quadrant, with an average value of $135~e^{-}/\mathrm{gr}/\mathrm{day}$. The lower $x$-axis shows the first ten days of measurement, while the upper $x$-axis indicates the corresponding iso-detection angles. Vertical error bars denote the statistical uncertainty of each image, given by the square root of the counts in the image, while horizontal error bars indicate the estimated average rate over the associated $6$-hour exposure (note that the $6$-hour exposures were not taken sequentially).
}
  \label{fig:data_VS_model}
\end{figure}

Any finite dataset will naturally exhibit statistical fluctuations around an ideal Poisson distribution. To quantify the expected range of such fluctuations under the null hypothesis, we generate an ensemble of Monte Carlo datasets with no injected signal assuming purely Poisson statistics. For each simulated dataset, we perform multiple bootstrap resamplings of the timestamps. For each resampled instance, we compute the $p$-value. We then average the $p$-values obtained from these bootstrap samples to assign a single mean $p$-value to that dataset. Repeating this procedure over the full ensemble of simulated datasets yields a distribution of mean $p$-values, that is used to define a $\pm2\sigma_p$-wide band within which the data are consistent with Poisson fluctuations, which is referred to as the Poisson-consistent band. A detailed explanation of the bootstrap simulation is provided in the Supplemental Materials.

\begin{figure*}[]
\centering
\includegraphics[width=0.99\linewidth]{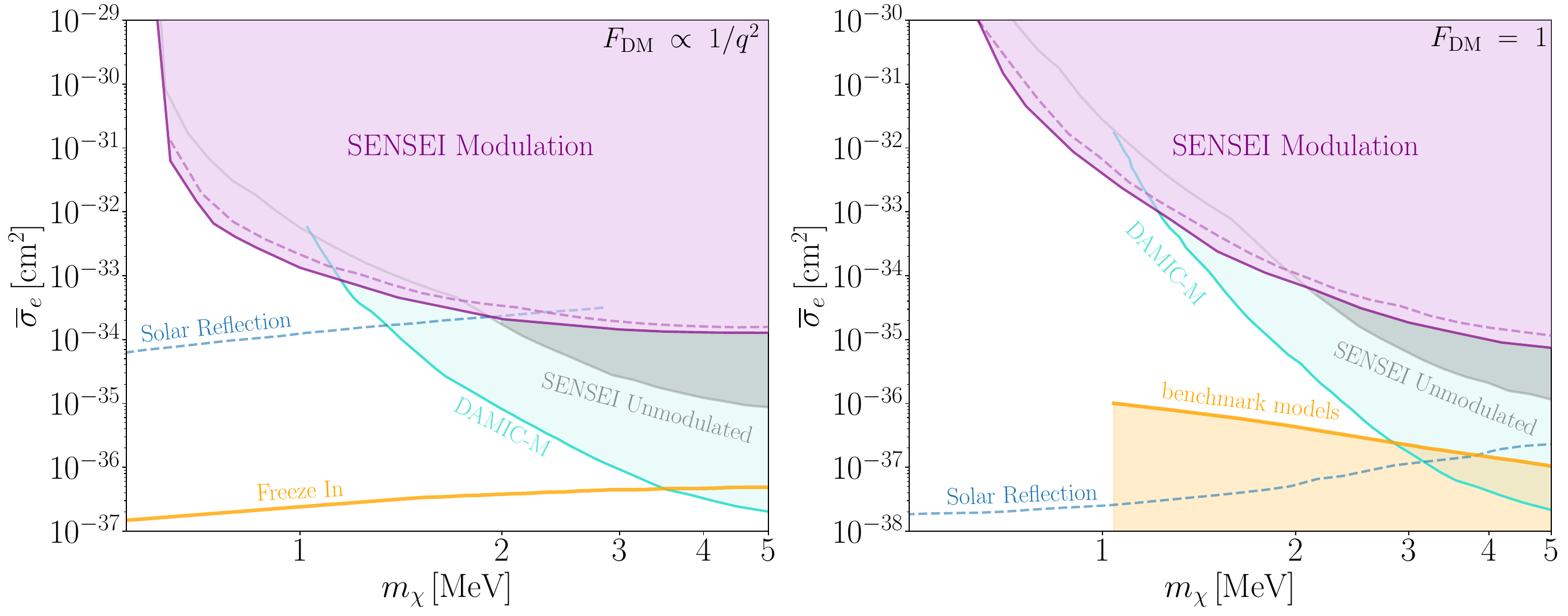}
\caption{ 90\% C.L. upper-limits on the DM electron scattering cross-section for a light ({\bf left}) and heavy ({\bf right}) dark-photon mediator, calculated with {\tt DaMaSCUS} and {\tt QCDark}. {\bf Solid purple line and shaded region} correspond to the constraints applied directly to the model referenced in Eq.~\eqref{eq:mod_dep_sig}, while the {\bf dashed line} 
indicates the 90\% C.L.~upper-limit obtained by recasting our constraint in Eq.~\eqref{eq:a1limit} on the model-independent modulation amplitude $a_1$ of Eq.~\eqref{eq:mod_ind_signal}; the latter was calculated by fitting the mass and $\bar{\sigma}_{e}$ that generates the same daily modulation amplitude $a_1$. The {\bf blue dashed curve} shows the constraint from solar‐reflected halo DM, assuming a dark‐photon mediator~\cite{Emken:2024nox,An:2021qdl, aprile2025search}. The {\bf orange solid curve} indicates the canonical freeze-in benchmark target for the light-mediator model~\cite{Essig:2011nj,Essig:2015cda,Chu:2011be,Dvorkin:2019zdi} and freeze-out region for the heavy mediator~\cite{Boehm:2003hm,Essig:2011nj,Essig:2015cda,Essig:2022dfa,Boehm:2003hm,Lin:2011gj,Izaguirre:2015yja,Hochberg:2014dra,Kuflik:2017iqs,DAgnolo:2019zkf,Chu:2011be,Dvorkin:2019zdi}. The {\bf solid gray line and shaded region} denote the existing SENSEI limits~\cite{senseisnolab2023,sensei1epaper}. The {\bf turquoise region}
show the DAMIC-M limit from~\cite{aggarwal2025probing}; we show the DAMIC-M daily modulation constraint from~\cite{DAMIC-M:2023hgj} in the Supplemental Materials, as it is based on {\tt QEDark} without screening.
}
\label{fig:results_qcdark_main}
\end{figure*}

We then apply the same bootstrap procedure to the observed dataset to compute its mean $p$-value. If this value lies within the Poisson-consistency band, the data are statistically compatible with the Poisson assumption of a time-independent background within our statistical treatment method. If, however, the mean $p$-value deviates significantly—i.e., by more than $2\sigma_p$—this would indicate a statistically significant departure from Poisson behavior. Such a deviation could correspond to a background fluctuation that mimics a signal-like effect large enough to substantially alter our exclusion limits. In that case, it is necessary to investigate the origin of the non-Poissonian background, whether due to unknown systematics or genuine time-dependent effects, before interpreting the result.

\section{Results}

For the signal models defined in Eqs.~\eqref{eq:mod_ind_signal} and \eqref{eq:model_dep_model}, the fits yield $\text{$p$-value} > 0.001$, indicating no statistically significant evidence for a DM signal at any mass. Results from the bootstrap test are presented in Fig.~\ref{fig: boot_simulation}. The mean $p$-value and statistical error of the observed data after performing the bootstrap procedure for a model-dependent signal (Eq.~\eqref{eq:mod_dep_sig}) is $0.52 \pm 0.02$ (blue band in Fig.~\ref{fig: boot_simulation}), while for a model-independent signal (Eq.~\eqref{eq:mod_ind_signal}) the $p$-value is $0.46 \pm 0.01$ (yellow band in Fig.~\ref{fig: boot_simulation}). In both cases, the mean $p$-values lie within the Poisson-consistency band. The figure also shows simulations using normal-ordered timestamps (solid lines) and bootstrap-resampled timestamps (dashed lines) for both signals, with each data point for the line obtained by calculating the mean of 500 $p$-values computed for simulated datasets at a given signal-to-noise ratio. 

Applying the procedure outlined in the Statistical Treatment section, we obtain a 90\%~C.L. upper limit on the model-independent modulation amplitude of
\begin{equation}
    a_1 \leq 6.8~\mathrm{e}^{-}\,\mathrm{g}^{-1}\,\mathrm{day}^{-1}\,.
    \label{eq:a1limit}
\end{equation}

For the model-dependent analysis, Fig.~\ref{fig:data_VS_model} compares the least noisy CCD quadrant of the data (shown in black), after subtraction of the best-fit background, with the expected DM signal from Eq.~\eqref{eq:model_dep_model} (after subtracting the mean DM signal rate). We assume a DM particle mass of $1$~MeV and a cross-section $\bar{\sigma}_{e}=5\times10^{-32},\mathrm{cm^{2}}$, interacting via a heavy dark-photon mediator. This quadrant dominates the resulting constraint on the model. 

Fig.~\ref{fig:results_qcdark_main} shows the 90\% C.L.\ exclusion limits for light (left panel) and heavy (right panel) mediators, computed using {\tt DaMaSCUS} and {\tt QCDark}. The solid purple line and shaded region represent the exclusion limits derived in this work. For comparison, previous constraints from SENSEI~\cite{senseisnolab2023, sensei_single_e_characterization} and DAMIC-M \cite{aggarwal2025probing} are also included. The daily-modulation search limit by DAMIC-M~\cite{DAMIC-M:2023hgj} is included only in the Supplemental Materials, since it was calculated with {\tt QEDark} without screening, and can thus not be directly compared.
This analysis significantly improves upon previous direct detection limits in the region dominated by the single-electron (1\e) channel, particularly for DM masses in the range of approximately $0.6-1$~MeV. The dashed purple lines correspond to exclusion limits obtained by identifying the values of $\overline{\sigma}_e$ and $m_\chi$ that saturate the model-independent limit from Eq.~\eqref{eq:a1limit}. 
As expected, these recasted limits are slightly weaker than those obtained from the full model-dependent analysis, since the model-independent constraint does not incorporate the full spectral information encoded in the specific model from Eq.~\eqref{eq:mod_dep_sig}.

In summary, we have presented both model~independent and model~dependent searches for daily modulation in the single-electron event rate measured with Skipper-CCDs at Fermilab. This approach enabled a significant improvement---by approximately an order of magnitude---in the constraints on halo DM scattering for DM masses below 1~MeV, relative to previous results, although the bounds are still weaker than those obtained from solar reflected DM. The sensitivity can be enhanced further by considering datasets with larger exposures. 

\section{Acknowledgements}
We are grateful for the support of the Heising-Simons Foundation under Grant No.~79921.
The CCD development work was supported in part by the Director, Office of Science, of the DOE under No.~DE-AC02-05CH11231. RE acknowledges support from DOE Grant DE-SC0025309 and Simons Investigator in Physics Award~MPS-SIP-00010469. 
TV is supported, in part, by the Israel Science Foundation (grant No. 1862/21), and by the NSF-BSF (grant No. 2021780). RE and TV acknowledge support from the Binational Science Foundation (grant No. 2020220).
IB is grateful for the support of the Alexander Zaks Scholarship, The Buchmann Scholarship, and the Azrieli Foundation.
AD, YW are supported by HSF and URA.
This document was prepared by the SENSEI collaboration using the resources of the Fermi National Accelerator Laboratory (Fermilab), a U.S. Department of Energy, Office of Science, Office of High Energy Physics HEP User Facility. Fermilab is managed by FermiForward Discovery Group, LLC, acting under Contract No. 89243024CSC000002. The U.S. Government retains and the publisher, by accepting the article for publication, acknowledges that the U.S. Government retains a non-exclusive, paid-up, irrevocable, world-wide license to publish or reproduce the published form of this manuscript, or allow others to do so, for U.S. Government purposes.

\section{Supplemental Materials}
In these Supplemental Materials, we present additional details of the analysis discussed in the main paper. 

\subsection{DM Search and Limit Calculations and Description of Likelihood Fitting}

The likelihood function for each image, $i=1,...,141$, and quadrant $q=1,...,8$, is modeled using the Poisson probability mass function:
\begin{gather}
    L_{i,q}(\mu, b_q \mid n_i^q) = \frac{[\lambda_{i,q}(\mu, b_q)]^{n^q_i}}{n^q_i !} \exp{[-\lambda_{i,q}(\mu, b_q)]}\,,
\label{eq:likelihood_function_single}
\end{gather}
where we define $\lambda_{i,q}(\mu, b_q)=(b_q+S(\mu))\epsilon_{i,q}$ to be the expected number of events after applying a cut efficiency of $\epsilon_{i,q}$ in image $i$ and quadrant $q$, $b_q$ is the number of dark-current events, and $S(\mu)$ is the number of DM signal events, given a model that depends on the set of parameters $\mu$. This accounts for the specific exposure time of each image and quadrant, as well as the selection efficiencies. The parameter $n_i^q$ is the  number of 1\e events in the corresponding image and quadrant.
The joint likelihood function is then the product over all images, 
\begin{equation}
    L(\mu, \bs{b} \mid \bs{n}) = \prod_{i, q} L_{i,q}(\mu, b_q \mid {n^i_q}) \,,
    \label{eq:joint_likelihood}
\end{equation}
where we use boldface notation to represent vector or matrix variables: $\bs{b} = \{b_q\}_q$, $\bs{n}=\{n^i_q\}_{i,q}$.

In what follows, we consider two hypotheses: the null hypothesis, $S(\mu=0)=0$,  corresponding to the background-only (no signal) case, and the alternative, $\mu\neq 0$, which corresponds to a non-vanishing signal contribution.  We define the test-statistic
\begin{equation}
     t_{\mu}(\bs{n}) = -2\ln\frac{L(\mu;  {\hat{\hat{\bs{b}}}}\mid \bs{n})}{L(\hat{\mu};{\hat{\bs{b}}\mid \bs{n})}} \,,
\label{eq:test-statistics}
\end{equation}
where $\hat{\hat{{\bs{b}}}}$ denotes the value of $\bs{b}$ that maximizes $L$ for the specified value of $\mu$, while $\hat{\mu}$ and
$\hat{\bs{b}}$ are the values that maximize $L$. 
Given a measured dataset, the corresponding $p$-value is then given by,  
\begin{equation}
p\text{-value}(t_{\mu}) = \int_{t_\mu^{\text{obs}}}^{\infty} f(t_{\mu}|\mu) \, dt_{\mu} \,,
\label{eq:p-value}
\end{equation}
where $f(t_{\mu}|\mu)$ is the probability density function (PDF) of the test-statistic $t_{\mu}$. 

For a discovery search, we consider the specific test-statistic $t_{\mu=0}(\bs{n})$, which evaluates the statistical validity of the null hypothesis. To derive its PDF, we use a Monte Carlo (MC) simulation to generate the distribution of the test-statistic, $f(t_{\mu}|\mu)$.
The background rate $b_q$ for each quadrant is estimated from real data by calculating the average event rate per quadrant, based on all measured images $n^q_i$.
Using this rate as the mean, we generate a Poisson-distributed background-only model for each quadrant.
To simulate the full dataset, we generate the total number of images (141) based on the background-only Poisson expectation $\lambda_{i,q}(\mu=0, b_q)$. Summing the resulting Poisson distributions gives our simulated dataset $\{\bs{n}(\mu=0)\}_{\rm sim}$, which provide a sample of corresponding values for the test-statistic $\{t_{\mu=0}\}_{\rm sim}$.
By repeating this procedure many times, we extract the corresponding PDF. 
Evidence for a signal can then be declared if the $p$-value for the measured test-statistic falls below 0.01.  Otherwise, we conclude that no signal is observed in the data.

To set a limit on a DM model, we now use the more general test statistic defined in Eq.~\eqref{eq:test-statistics}. For the MC simulation, we repeat the procedure described above for each distinct value of $\mu$, using $\lambda_{i,q}(\mu, \hat{\hat{\bs{b}}})$, where again $\hat{\hat{\bs{b}}}$ is the likelihood-maximizing value for the real data and the specified $\mu$. For each simulated dataset, the test statistic is computed separately, re-estimating $\hat{\mu}$, $\hat{\bs{b}}$, and $\hat{\hat{\bs{b}}}$ in each iteration. The resulting distribution $f(t_{\mu} | \mu)$ is then obtained and used to derive the 90\%  confidence level (CL) limit 
 by determining the values of $\mu$ for which $\text{$p$-value}(t_{\mu}) = 0.1$.

\subsection{Impact of Non-Poisson Backgrounds on the Results}

\begin{figure}[h]
  \centering
\includegraphics[width=1\linewidth]{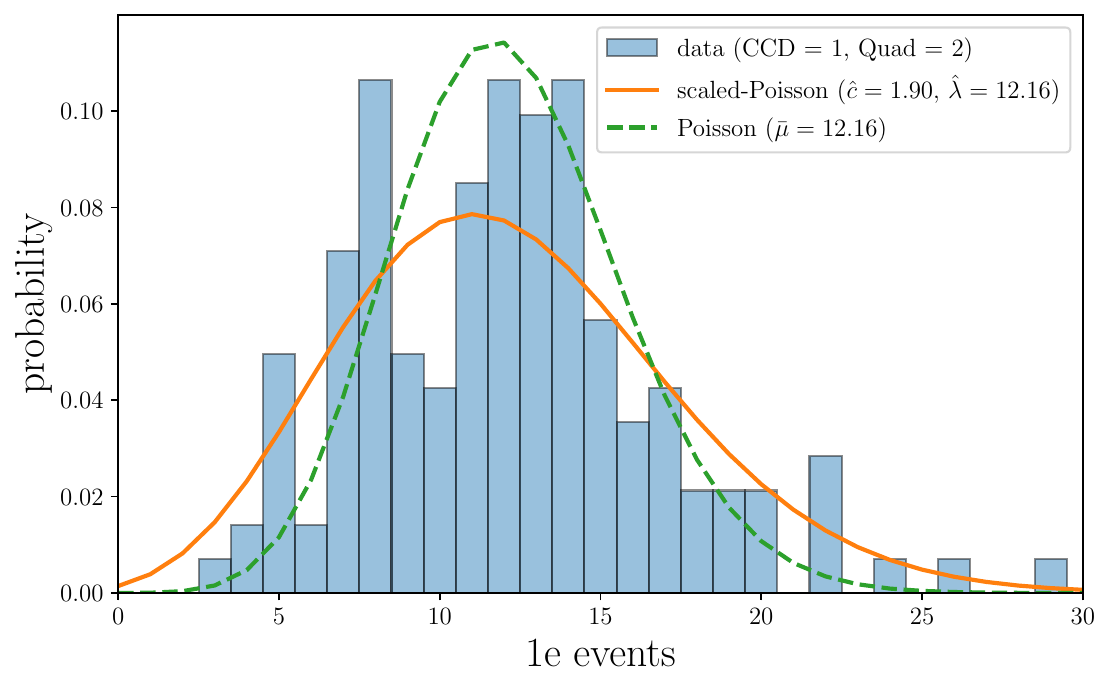}
  \caption{Measured single-electron event spectrum from CCD~1, quadrant~2 (blue bins), together with the best-fit \textit{scaled-Poisson} background model (orange curve, $\hat{c} = 1.90$, $\hat{\lambda} = 12.16$) and the corresponding standard Poisson distribution (green dashed curve, $\hat{\mu} = 12.16$). The comparison shows that the scaled-Poisson model captures the broader tail of the data more accurately than the Poisson model.
}
  \label{fig:Scaled_Pois_distirbution_Pois_Comp}
\end{figure}

\begin{figure*}[t]
  \centering
\includegraphics[width=1\linewidth]{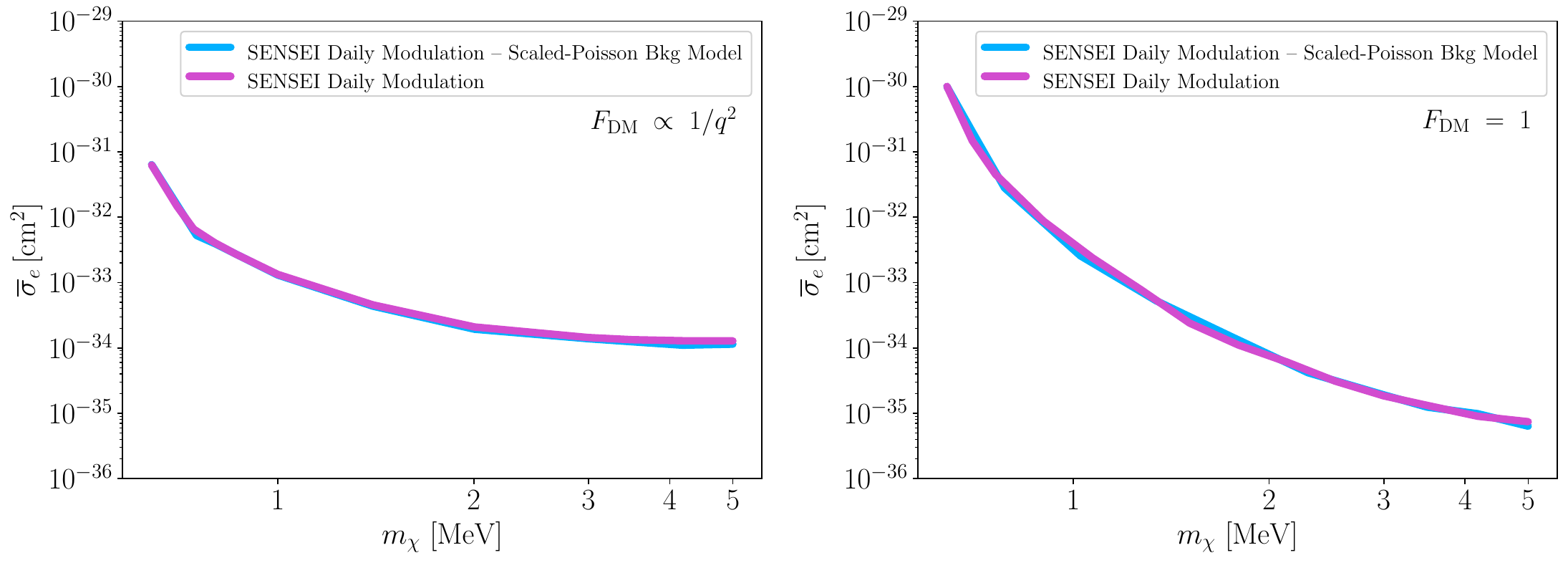}
  \caption{Comparison of our main results obtained with the scaled-Poisson background model (blue) and the Poisson background model (purple; same as Fig.~\ref{fig:results_qcdark_main}), shown for a light mediator (\textbf{left}) and a heavy mediator (\textbf{right}). The two curves agree within $10\%$, indicating that the analysis is not affected by our non-Poissonian background deviations.
}
  \label{fig:Scaled_Pois_test}
\end{figure*}

When examining the data, we observed a deviation from Poisson statistics, characterized primarily by a variance larger than that expected under a Poisson process. To assess the potential impact of such excess variance on our results, we constructed an alternative background model, which we term the \textit{Scaled-Poisson}, defined as a discrete distribution on $k$ with probability mass function
\begin{equation}
p(k\,|\,\mu,c)=\frac{e^{-\mu/c}\,(\mu/c)^{k/c}/\Gamma(k/c+1)}{\sum_{m=0}^{\infty} e^{-\mu/c}\,(\mu/c)^{m/c}/\Gamma(m/c+1)}\ ,
\label{eq:scaled_poisson_dist}
\end{equation}
where $\mu$ denotes the mean rate parameter (here $\mu\equiv\lambda_{i,\,q}$) and $c>0$ is a scale parameter, with $c=1$ corresponding to the Poisson distribution. 
While this choice of model is neither unique nor physically motivated, it provides a convenient framework to evaluate the effect of accommodating excess variance in the background on our results. Fitting this model to the data yielded a best-fit scale factor of $c = 1.9$. Comparison of the \textit{Scaled-Poisson} distribution~\eqref{eq:scaled_poisson_dist} and Poisson distribution with data from the second quadrant of the first CCD is shown in Fig.~\ref{fig:Scaled_Pois_distirbution_Pois_Comp}.
Fig.~\ref{fig:Scaled_Pois_test} compares the results of the main analysis (Fig.~\ref{fig:results_qcdark_main}) with those obtained under the Scaled-Poisson background model. The two limit curves agree within $10\%$. This level of agreement is consistent with the outcome of the bootstrap test, which indicated that the time-independent background is compatible with a Poisson distribution within our statistical framework.

\subsection{Bootstrap Simulation Tests}
\label{sec:Bootstarp_Method}
To test that bootstrapping the data eliminates a DM signal, we employ the Monte Carlo simulation described in the Statistical Treatment section of the main paper. For a given signal strength $\mu$, we simulate the $\lambda_{i,q}(\mu,b_q)$, and a quadrant-dependent number of background events that is estimated from the time‐averaged rate in the observed data, $\langle n_q^i \rangle_i$, and the mean number of signal events, $\langle S(\mu)\rangle$, via $b_q = \langle n_q^i \rangle_i - \langle S(\mu)\rangle.$ This procedure yields a dataset whose total rate matches that of the observed data with a given signal $\mu$. Using this combined rate, we then generate a Poisson‐distributed dataset under the hypothesis that a signal is present with the same timestamps as the observed hidden data.
 Repeating this procedure many times for each $\mu$ and calculating $p\text{-value}(t_{\mu=0})$ using Eq.~\eqref{eq:p-value} yields a set of $p\text{-value}$s from which we extract the mean. Then we bootstrap the same datasets and repeat the calculation of $p\text{-value}(t_{\mu=0})$ to extract the mean. This yields two curves of the mean $p$-value versus $\mu$: one for the original time-ordered data %timestamps 
 and one for the bootstrapped time-ordered data. This process is done for the two signals Eqns.~\eqref{eq:mod_ind_signal} and \eqref{eq:mod_dep_sig} while increasing the signal $\mu$.
 For the model-dependent signal, we define the signal-to-noise (SNR) as the sum of the mean DM signal and the daily modulation rate over the mean background, averaged over all quadrants. In the case of the model-independent signal, the SNR is obtained from the daily modulation amplitude $a_1$ [Eq.~\eqref{eq:mod_ind_signal}] over the mean background across all quadrants.

\begin{figure*}[t!]
\centering
\includegraphics[width=0.99\linewidth]{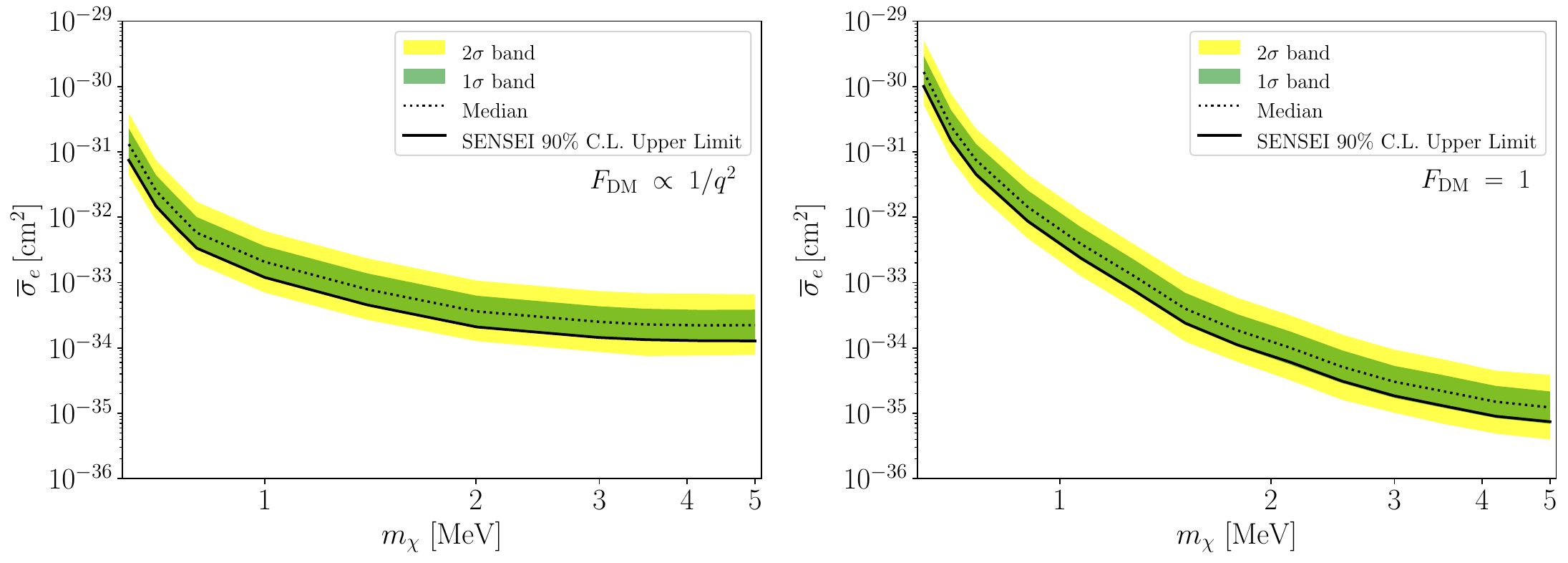}
\caption{Expected sensitivity bands for DM-electron interactions via light (\textbf{left}) and heavy (\textbf{right}) mediator calculated with {\tt DaMaSCUS} and {\tt QCDark}. Green and yellow regions stand for the 68.3\% ($1\sigma$) and 95.5\% ($2\sigma$) of the expected band distribution, respectively. The black dotted line shows the median of the expected distribution, and the solid black line is the upper limit from Fig. \ref{fig:results_qcdark_main}.}
\label{fig: Expected_bands_QCDark}
\end{figure*}

\begin{figure*}[t]
\centering
\begin{minipage}{0.495\textwidth}
  \centering
  \includegraphics[width=\linewidth]{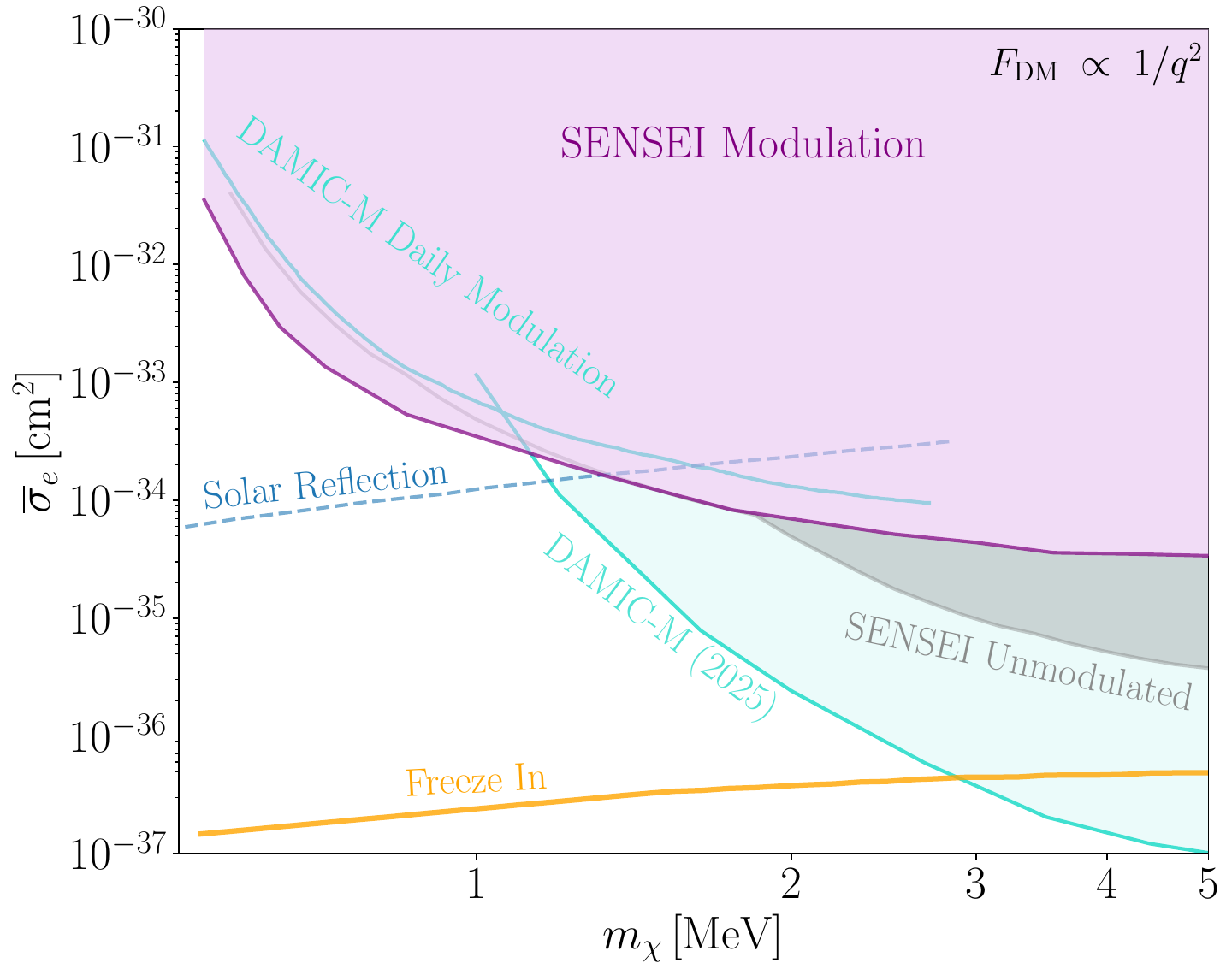}
\end{minipage}%
\hfill%
\begin{minipage}{0.495\textwidth}
  \centering
  \includegraphics[width=\linewidth]{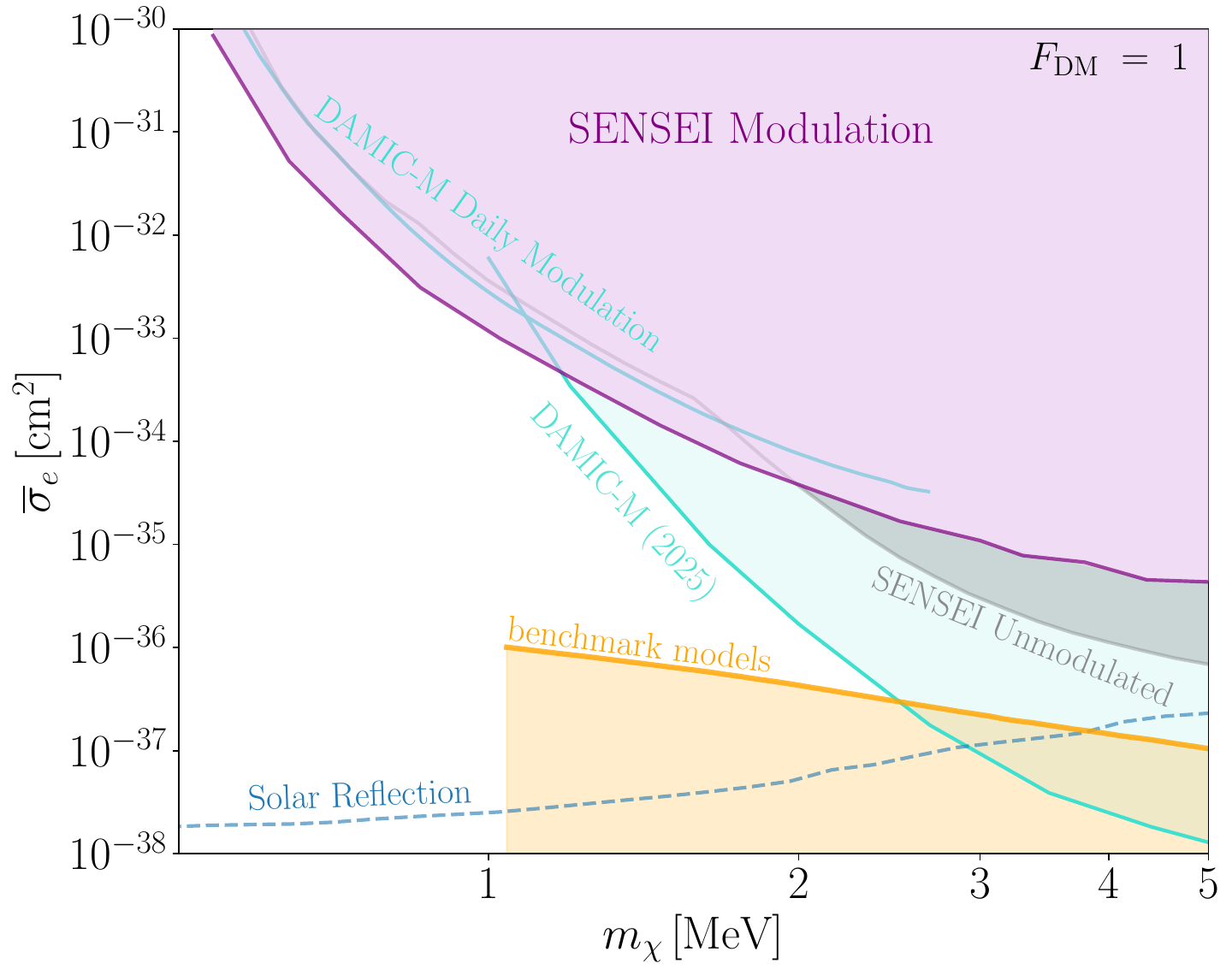}
\end{minipage}
\caption{
90\% C.L. upper-limits on the DM electron scattering cross-section for light ({\bf left}) and heavy ({\bf right}) dark-photon mediators, calculated with {\tt Verne}~\cite{Verne2017_Kavanagh} and {\tt QEDark}~\cite{Essig:2015cda} (without screening). {\bf Solid purple line and shaded region} denote the constraints on the model referenced in Eq.~\eqref{eq:model_dep_model}. The {\bf blue dashed curve} shows the constraint from solar‐reflected halo DM, assuming a dark‐photon mediator~\cite{Emken:2024nox,An:2021qdl, aprile2025search}. The {\bf orange solid curve} indicates the canonical freeze-in benchmark target for the light mediator model~\cite{Essig:2011nj,Essig:2015cda,Chu:2011be,Dvorkin:2019zdi} and freeze-out region for the heavy mediator~\cite{Boehm:2003hm,Essig:2011nj,Essig:2015cda,Essig:2022dfa,Boehm:2003hm,Lin:2011gj,Izaguirre:2015yja,Hochberg:2014dra,Kuflik:2017iqs,DAgnolo:2019zkf,Chu:2011be,Dvorkin:2019zdi}. The {\bf solid gray line and shaded region} corresponds to previous SENSEI limits 
~\cite{senseisnolab2023,sensei1epaper}. The {\bf turquoise line and region}
 are the DAMIC-M daily-modulation limit~\cite{DAMIC-M:2023hgj} and limits from~\cite{aggarwal2025probing} by DAMIC-M.}

\label{fig: results_qedark_no_screen}
\end{figure*}

Fig.~\ref{fig: boot_simulation} presents the simulation results for 500 iterations of fits for each point. Yellow markers show the mean $p$-value for fits to the model-independent signal (Eq. \eqref{eq:mod_ind_signal}), while the blue ones are for the model-dependent signal (Eq. \eqref{eq:mod_dep_sig}). Solid lines correspond to results with normal-ordered timestamps, dot-dashed lines to those obtained after bootstrapping the timestamps of the datasets, and the dashed lines correspond to the mean $p$-value bootstrap calculation of the real data. At low SNR, all curves converge to the Poisson consistency band, as expected: when the signal is negligible, the datasets are dominated by Poisson background and thus yield consistent results with the no-signal hypothesis. In the presence of a signal, the mean $p$-value, $p\text{-value}(t_{\mu=0})$, decreases with SNR in the time-ordered simulated datasets, whereas it remains flat for the bootstrapped datasets. This demonstrate that bootstrapping removes signals up to $0.7$ and $0.55$ in SNR for model-dependent and model-independent signals, respectively.

\subsection{Expected Bands for Exclusion Limits}
In Fig.~\ref{fig: Expected_bands_QCDark}, we display the expected sensitivity bands, obtained from 10\,000 simulated datasets generated by the Monte Carlo simulation based on the background-only hypothesis described above. For each simulated dataset we compute a 90\% C.L.~upper limit, and the sensitivity bands are defined by the distribution of these limits. The green region and yellow region denotes the central 68.3\% ($1\sigma$) and 95.5\% ($2\sigma$) of the distribution, respectively. The upper limit on the observed data from Fig.~\ref{fig:results_qcdark_main} is shown in solid black and falls within the $2 \sigma$ sensitivity band.

\subsection{Additional Dark-Matter Exclusion Limits}

To compare our results with previous exclusion limits, we note that the search performed in~\cite{Arnquist_2024} uses {\tt QEDark}~\cite{Essig:2015cda} without the charge-screening function in~\cite{dreyer2023fullyabinitioallelectroncalculation}, as well as a different simulation software, {\tt Verne}, for calculating the distorted velocity distributions. 
{\tt Verne} models light DM particles as traveling in straight-line trajectories, either straight through the Earth or undergoing complete reflection upon a single scatter. This code is efficient and fast, but as it assumes only one scatter, it is expected to lose accuracy if the DM-nucleon cross-section is large enough such that DM would undergo multiple scatters in the Earth. 

Fig.~\ref{fig: results_qedark_no_screen} shows the constraints on DM scattering through a light (\textbf{left}) and heavy (\textbf{right}) mediator calculated with {\tt QEDark}~\cite{Essig:2015cda,QEdark} and {\tt Verne}~\cite{Verne2017_Kavanagh} without charge screening, in order to compare our results with other relevant constraints in the literature. These plots demonstrate that this work's constraints are more stringent for the 1e channel compared to previous constraints in the literature.

\bibliographystyle{apsrev4-1}
\bibliography{biblio}

\end{document}

%% file: authors.tex
\author{Itay M. Bloch}
\affiliation{Berkeley Center for Theoretical Physics, University of California, Berkeley, CA 94720, U.S.A.}
\affiliation{Theoretical Physics Group, Lawrence Berkeley National Laboratory, Berkeley, CA 94720, U.S.A.}

\author{Ana M. Botti}
\affiliation{\normalsize\it 
Fermi National Accelerator Laboratory, PO Box 500, Batavia IL, 60510, USA}
\affiliation{\normalsize\it Kavli Institute for Cosmological Physics, University of Chicago, Chicago, IL 60637, USA}

\author{Mariano Cababie}
\affiliation{\normalsize\it
Institut f\"ur Hochenergiephysik der \"Osterreichischen Akademie der Wissenschaften, 1050 Wien - Austria}
\affiliation{\normalsize\it
Atominstitut, Technische Universit\"at Wien, 1020 Wien - Austria}
\affiliation{\normalsize\it 
Fermi National Accelerator Laboratory, PO Box 500, Batavia IL, 60510, USA}

\author{Gustavo Cancelo}
\affiliation{\normalsize\it 
Fermi National Accelerator Laboratory, PO Box 500, Batavia IL, 60510, USA}

\author{Brenda A. Cervantes-Vergara}
\affiliation{\normalsize\it 
Universidad Nacional Aut\'onoma de M\'exico, Ciudad de M\'exico, M\'exico}

% Removed by JT Oct-23-2024
%\author{Michael Crisler}
%\affiliation{\normalsize\it 
%Fermi National Accelerator Laboratory, PO Box 500, Batavia IL, 60510, USA}

\author{Miguel Daal}
\affiliation{\normalsize\it 
 School of Physics and Astronomy, 
 Tel-Aviv University, Tel-Aviv 69978, Israel}

\author{Ansh Desai}
\affiliation{\normalsize\it 
Department of Physics and Institute for Fundamental Science, University of Oregon, Eugene, Oregon 97403, USA}

\author{Alex Drlica-Wagner}
\affiliation{\normalsize\it 
Fermi National Accelerator Laboratory, PO Box 500, Batavia IL, 60510, USA}
\affiliation{\normalsize\it Kavli Institute for Cosmological Physics, University of Chicago, Chicago, IL 60637, USA}
\affiliation{\normalsize\it  Department of Astronomy and Astrophysics, University of Chicago, Chicago IL 60637, USA}

 \author{Rouven Essig}
\affiliation{\normalsize\it 
C.N.~Yang Institute for Theoretical Physics, Stony Brook University, Stony Brook, NY 11794, USA}

 \author{Juan Estrada}
\affiliation{\normalsize\it 
Fermi National Accelerator Laboratory, PO Box 500, Batavia IL, 60510, USA}

\author{Erez Etzion}
\affiliation{\normalsize\it 
 School of Physics and Astronomy, 
 Tel-Aviv University, Tel-Aviv 69978, Israel}

\author{Guillermo Fernandez Moroni}
\affiliation{\normalsize\it 
Fermi National Accelerator Laboratory, PO Box 500, Batavia IL, 60510, USA}

\author{Stephen E. Holland}
\affiliation{\normalsize\it 
Lawrence Berkeley National Laboratory, One Cyclotron Road, Berkeley, California 94720, USA}

\author{Jonathan Kehat}
\affiliation{\normalsize\it 
 School of Physics and Astronomy, 
 Tel-Aviv University, Tel-Aviv 69978, Israel}

\author{Ian Lawson}
\affiliation{\normalsize\it SNOLAB, Lively, ON P3Y 1N2, Canada}

\author{Steffon Luoma}
\affiliation{\normalsize\it SNOLAB, Lively, ON P3Y 1N2, Canada}

 \author{Aviv Orly}
\affiliation{\normalsize\it 
 School of Physics and Astronomy, 
 Tel-Aviv University, Tel-Aviv 69978, Israel}

\author{Santiago E. Perez}
\affiliation{\normalsize\it 
Fermi National Accelerator Laboratory, PO Box 500, Batavia IL, 60510, USA}
\affiliation{\normalsize\it 
Universidad de Buenos Aires, Facultad de Ciencias Exactas y Naturales, Departamento de Física, Buenos Aires, Argentina}
\affiliation{\normalsize\it 
CONICET - Universidad de Buenos Aires, Instituto de Física de Buenos Aires (IFIBA). Buenos Aires, Argentina}

\author{Dario Rodrigues}
\affiliation{\normalsize\it 
Universidad de Buenos Aires, Facultad de Ciencias Exactas y Naturales, Departamento de Física, Buenos Aires, Argentina}
\affiliation{\normalsize\it 
CONICET - Universidad de Buenos Aires, Instituto de Física de Buenos Aires (IFIBA). Buenos Aires, Argentina}

\author{Nathan A. Saffold}
\affiliation{\normalsize\it 
Fermi National Accelerator Laboratory, PO Box 500, Batavia IL, 60510, USA}

\author{Silvia Scorza}
\affiliation{\normalsize\it 
Univ. Grenoble Alpes, CNRS, Grenoble INP, LPSC-IN2P3, Grenoble, 38000, France}

 \author{Miguel Sofo-Haro}
\affiliation{\normalsize\it 
Fermi National Accelerator Laboratory, PO Box 500, Batavia IL, 60510, USA}
\affiliation{Universidad Nacional de C\'ordoba, CNEA/CONICET, C\'ordoba, Argentina}

% Removed by JT Oct-23-2024
%\author{Leandro Stefanazzi}
%\affiliation{\normalsize\it 
%Fermi National Accelerator Laboratory, PO Box 500, Batavia IL, 60510, USA}

 \author{Kelly Stifter}
\affiliation{\normalsize\it 
Fermi National Accelerator Laboratory, PO Box 500, Batavia IL, 60510, USA}

\author{Javier Tiffenberg}
\affiliation{\normalsize\it 
Fermi National Accelerator Laboratory, PO Box 500, Batavia IL, 60510, USA}

\author{Sho Uemura}
\affiliation{\normalsize\it 
Fermi National Accelerator Laboratory, PO Box 500, Batavia IL, 60510, USA}

\author{Edgar Marrufo Villalpando}
\affiliation{\normalsize\it Kavli Institute for Cosmological Physics, University of Chicago, Chicago, IL 60637, USA}

\author{Tomer Volansky}
\affiliation{\normalsize\it 
 School of Physics and Astronomy,   
 Tel-Aviv University, Tel-Aviv 69978, Israel}

% Added by JT Oct-23-2024
\author{Federico Winkel}
\affiliation{\normalsize\it 
Universidad de Buenos Aires, Facultad de Ciencias Exactas y Naturales, Departamento de Física, Buenos Aires, Argentina}
\affiliation{\normalsize\it 
CONICET - Universidad de Buenos Aires, Instituto de Física de Buenos Aires (IFIBA). Buenos Aires, Argentina}

\author{Yikai Wu}
\affiliation{\normalsize\it 
C.N.~Yang Institute for Theoretical Physics, Stony Brook University, Stony Brook, NY 11794, USA}
\affiliation{\normalsize\it 
Department of Physics and Astronomy, Stony Brook University, Stony Brook, NY 11794, USA} 

\author{Tien-Tien Yu}
\affiliation{\normalsize\it 
Department of Physics and Institute for Fundamental Science, University of Oregon, Eugene, Oregon 97403, USA}

\collaboration{The SENSEI Collaboration }

\author{Xavier Bertou}
\affiliation{\normalsize\it
CNRS/IN2P3, IJCLab, Université Paris-Saclay, Orsay, France}

% sho's notes:
% people who were borderline for 2023 paper (Silvia, Leo, Edgar)
% people who have left (Yaron, Kelly)
% people whose status I don't know (Prakruth, Mariano)